# Absence of Ferromagnetism in Mn-doped Tetragonal Zirconia

*S. K. Srivastava[*,1], P. Lejay[1], B. Barbara[1], O. Boisron[2], S. Pailhès[2,3] and G. Bouzerar[1,4]*

[1]Institut Néel, CNRS Grenoble and Université Joseph Fourier, 38042 Grenoble Cedex 9, France

[2]University Lyon 1 and CNRS, 69622 Villeurbanne, France

[3]Laboratoire Léon Brillouin (LLB), CEA/CNRS Saclay, 91191 Gif-sur-Yvette, France

[4]Bremen School of Engineering and Science, Jacobs University, Campus Ring 1, D-28759 Bremen, Germany

## Abstract

In a recent letter, it has been predicted within first principle studies that Mn-doped $ZrO_2$ compounds could be good candidate for spintronics application because expected to exhibit ferromagnetism far beyond room temperature. Our purpose is to address this issue experimentally for Mn-doped *tetragonal* zirconia. We have prepared polycrystalline samples of $Y_{0.15}(Zr_{0.85-y}Mn_y)O_2$ (y=0, 0.05, 0.10, 0.15 & 0.20) by using standard *solid state method at equilibrium*. The obtained samples were carefully characterized by using x-ray diffraction, scanning electron microscopy, elemental color mapping, X-ray photoemission spectroscopy and magnetization measurements. From the detailed structural analyses, we have observed that the 5% Mn doped compound crystallized into two symmetries (dominating tetragonal & monoclinic), whereas higher Mn doped compounds are found to be in the tetragonal symmetry only. The spectral splitting of the Mn 3*s* core-level x-ray photoelectron spectra confirms that Mn ions are in the $Mn^{3+}$ oxidation state and indicate a local magnetic moment of about 4.5 $\mu_B$/Mn. Magnetic measurements showed that compounds up to 10% of Mn doping are paramagnetic with antiferromagnetic interactions. However, higher Mn doped compound exhibits local ferrimagnetic ordering. Thus, no ferromagnetism has been observed for all Mn-doped *tetragonal* $ZrO_2$ samples.

---

[*] **E-mail:** sandeep.srivastava@grenoble.cnrs.fr

# I. INTRODUCTION

In recent years, spintronics has become one of the most important fields of research as demonstrated by the huge existing literature. A frenetic race which involves both fundamental research scientists and industrial partners has been recently triggered to discover and elaborate a new family of materials for spintronics: the ferromagnetic semi-conductors. The interest in such materials consists in the use of new devices where spin degrees of freedom carry information in order to reduce electrical consumption, allow non-volatility, store and manipulate data, beyond room temperature. Towards this end, one approach is to introduce magnetic dopants into large gap semi-conductor solids, and hope that they not only remain magnetically active but also couple with the electronic states of the solid. Considerable success has been achieved in inducing room temperature ferromagnetism by substituting transition elements in semiconductor host such as ZnO, $TiO_2$, $SnO_2$ [1-5].

Among the oxide materials, $ZrO_2$ is a promising material with high dielectric constant and ionic conductivity and, it is used in industry with applications in catalysis and in solid-oxide fuel cells [6]. It can be crystallized in a monoclinic phase (at ambient temperature) or cubic or tetragonal phases (at very high temperature) [7]. One of the major challenges is therefore to control and tune the crystallographic phase of 3d ions doped Zirconia. The tetragonal and cubic phases can be stabilized by the addition of another cation such as $Ca^{2+}$ or $Y^{3+}$ [8-9]. Recently it was predicted from ab initio electronic structure calculations that cubic $ZrO_2$ (zirconia) stabilized by Mn or doped with other transition elements such as Co and Fe should be ferromagnetic with an ordering temperature above 500 K [10]. Another calculation for both interstitial and substitutional Mn for monoclinic, tetragonal, and cubic zirconia with 25% Mn has been done, predicting the moment of around $3\mu_B$/Mn atom [11]. Moreover, theoretical calculation has predicted ferromagnetism in K-doped $ZrO_2$ also [12], while Cu [13] or Cr [10] or Ca [12] doped $ZrO_2$ result in paramagnetism, antiferromagnetic or nonmagnetic ground states respectively. Motivated by these promising theoretical predictions, few experimental works has been done to confirm the ferromagnetism in Mn-doped zirconia. Manganese-doped zirconia nano-crystals reported by Clavel *et al.* were found to be purely paramagnetic at room temperature [14]. Moreover, Mn and Fe-stabilized cubic zirconia nanoparticles with up to 35% and 40% $3d$-element content were also found to be paramagnetic at room temperature and at 5 K [15]. However, Zippel *et al.* have reported defect induced ferromagnetism at room temperature for both undoped and Mn-doped zirconia thin films with up to 20 at. % Mn [16].

The search for candidates with room temperature ferromagnetism is really intense but, in most cases, preparation of materials is not very well controlled. The inconsistence and controversy may arise from the poor characterization of the samples, including stoichiometry, homogeneity, phase segregation, etc. For thin films, the stoichiometry and phase purity could be difficult to establish and the metastable preparation conditions likely results in phase segregation. It is therefore imperative to prepare bulk materials *at equilibrium* conditions, which will intrinsically diminish the uncertainties and inaccuracies in characterization. Motivated by these interesting theoretical prediction for Mn-doped zirconia [10, 11], we have undertaken the experimental work on Mn-doped *tetragonal* zirconia to look the role of tetragonal symmetry on the magnetic property, if any. We have prepared a series of polycrystalline samples of $Y_x(Zr_{1-x-y}Mn_y)O_2$ for the values of x = 0-0.15 and y = 0- 0.15 with the intention to stabilize the tetragonal structure, and found that Y substitutions can stabilize a single phase of tetragonal structure containing Mn ions for x ~ 0.15.

## II. EXPERIMENTAL DETAILS

The polycrystalline compounds $Y_{0.15}(Zr_{0.85-y}Mn_y)O_2$ (y=0, 0.05, 0.10, 0.15 & 0.20) were prepared by standard solid state route method by using high-purity $ZrO_2$, $Y_2O_3$ and $MnO_2$ compounds. The final annealing in pellet form was carried out at $1550^0C$ for 70 hrs. Slow scan X-Ray diffraction patterns were collected by using Philips XRD machine with $CuK_\alpha$ radiation at room temperature. Recording of microstructure image and elemental analysis have been carried out by using Zeiss-Ultra Scanning Electron Microscope (SEM) equipped with energy dispersive spectrometer (EDS). The XPS spectra were recorded with a CLAM 4 vacuum generator (Al K line at 1486.6 eV), and the photoelectrons were collected at a pass energy of 20 eV in the fixed analyzer transmission mode, which gives a full width half maximum (FMHM) of just over one eV. Charge referencing was done against adventitious carbon (C 1s=284.6 eV) attributed to the CO contribution present in each sample. The Shirley-type background was subtracted from the recorded spectra and the curve fitting and deconvolution of overlapped peaks were done by nonlinear least-square fitting (gnuplot interface) with a Gauss–Lorentz curve. The surfaces for all the samples except for the reference $MnO_2$ powder were cleaned with argon ion sputter in order to improve the ratio signal to background. Tests of measurements performed on few samples (before and after the ions bombardment) convinced us that the samples are stable at sputtering contrary to $MnO_2$ which reduces to $Mn_2O_3$. Magnetization measurements as a function of magnetic field (H) and temperature (T) were carried out using SQUID magnetometer and high field magnetometer developed at Institut Neel, CNRS Grenoble, France.

## III. RESULTS AND DISCUSSIONS

The XRD patterns of these samples are shown in Figure 1. For the 5% Mn doped sample, all the diffraction peaks could be indexed to the tetragonal symmetry with weak supplementary peaks indexed to the monoclinic symmetry. However, the diffraction peaks of higher Mn compound (upto 15%) could be indexed to the tetragonal phase only without any parasitic phase. For 20% Mn doped compound, additional impurity phase indexed as $Mn_3O_4$ starts to appear in addition to main tetragonal phase. The lattice parameters, as shown in Figure 2 (e), obtained from the refinement of XRD patterns with Fullprof program [17], are found to decrease significantly with Mn doping, indicating solubility up to 15% Mn concentration. The decrease of lattice parameters upon Mn doping in $Zr_{0.85}Y_{0.15}O_2$ can be understood on the basis of doping of $Mn^{3+}$ ion (confirmed in the later section, with ionic radii of 0.645 Å) to the $Zr_{0.85}Y_{0.15}$ (average ionic radii 0.747 Å). The abrupt increase of lattice parameters for 20% Mn doped compound can be understood in terms of introduction of $Mn_3O_4$ impurity phase. The morphology of all the samples has been studied by recording scanning electron micrographs (SEM). The morphology of the samples up to 10% of Mn doping was found to be quite uniform whereas the 15% Mn doped compound starts to show grains boundary. The EDS analysis confirms that the composition of the prepared materials is comparable to the nominal starting one. The average cationic ratio for 15% of Mn doped compounds is found to be Mn: Y: Zr =15.37: 14.19: 70.44. To understand the appearance of grain boundary in 15% Mn doped compound, the elemental color mapping by EDS has been performed. The grain boundaries associated with it are found to have thickness of about 3 micrometer. The elemental color mapping by EDS (Figure 2) shows that the grain boundary region is richer in manganese and oxygen and this indicates that there are very tiny phases of Mn in 15 % Mn doped compound which could not be reflected in XRD.

The Mn 2p and 3s x-ray photoelectron (XP) spectra for 10, 15 and 20% Mn doped compounds are shown in the Figure 3 (a) and (b) respectively and the binding energies (BE) are listed in Table 1. For comparison, the corresponding Mn 2p and 3s spectra of $MnO_2$ powder used as starting material are also reported in the same figures. The Mn 2p spectra display two broad emissions lines for both $Mn2p_{3/2}$ and $Mn2p_{1/2}$, each followed by charge-transfer satellites at relative binding energy (BE) of about 4.5 eV. The resulting asymmetric shapes were then deconvoluted into four peaks. The BE of the Mn $2p_{(1/2, 3/2)}$ are given by the energy positions of the dominant peaks. The ratio of the intensities between the dominant peaks and their satellites are more than 2.5. The BE of Mn $2p_{3/2}$ in $MnO_2$ compound was found to be 642.6 eV, which is in agreement with other studies [19]. For $Y_{0.15}(Zr_{0.85-y}Mn_y)O_2$ compounds, the BE of the Mn $2p_{3/2}$ is found to be 640.6 eV when averaging between three compositions. This value is closer to BE measured in MnO (640.7eV for Mn $2p_{3/2}$) [20] than 641.8 eV found in $Mn_2O_3$ [20] indicating that the oxidation states of the Mn ions would be of +2. On the contrary, the analyses of the Mn 3s spectra reveal unambiguously an oxidation state of +3. In fact, the Mn 3s spectra exhibit two components which originate from the exchange interaction between the 3s core hole and open 3d shell and the splitting is therefore strongly dependent on the valence states of the Mn ions being of about 6.5eV for $Mn^{2+}$ in MnO, 5.5eV for $Mn^{3+}$ in $Mn_2O_3$ and 4.5eV for $Mn^{4+}$ in $MnO_2$ [21]. The splitting magnitude measured in the samples with the Mn contents y=0.1 and y=0.15 is close to 5.5 eV which corresponds to the expected splitting for $Mn^{+3}$. Note that the 3s splitting magnitude measured in our powder of $MnO_2$ was found to be 4.2eV which corresponds to the oxidation state of +4. Although, the 3s energy peaks positions in 20% Mn doped compound decrease abruptly toward the lower binding energy in comparison to y=0.1 and 0.15% Mn doped compounds, the 3s exchange splitting magnitude remains close to 5.5eV. Thus, from the Mn 3s exchange splitting, one can conclude that the oxidations states of Mn are close to +3 in these compounds. In systems, where the charge transfer satellites in the 2p spectra are small, *i.e.* small covalency, it has been shown that the magnitude of the Mn 3s splitting is proportional to the magnetic moment of the unpaired 3d electrons [22]. By using the linear calibration curve established by Kowalczyk [23], the magnitude of the 3s splitting measured in our samples corresponds to a local magnetic moment of ~4.5 $\mu_B$/Mn. Although the charge transfer peaks in the Mn 2p spectra can not be considered as being small, there is a good agreement between the magnetic moment deduced from the Mn 3s splitting and from the magnetic moment provided by the magnetization measurements. A slight tendency toward a decrease of the magnitude of the 3s splitting energy as function of Mn is also in agreement with the decrease of the effective paramagnetic moment extracted from the SQUID measurements discussed in the next section.

In order to characterize the XP properties of the $Y_{0.15}(Zr_{0.85-y}Mn_y)O_2$ samples and to compare with others type of zirconia, we also reported the XP spectra of the Zr and Y 3d(3/2 and 5/2) core levels in Figure 3 and the BE are listed in the table 1. For comparison, we report the energies of the Zr/Y 3d core levels measured in the tetragonal and cubic phase of the yttria-stabilized zirconia (YSZ) reported by other groups [24-26]. The Y/Zr core level energies and the chemical shifts are within an interval of 0.5eV similar with those measured in cubic and tetragonal YSZ [24-26]. It confirms the tetragonal symmetry, also deduced from the XRD analysis. When compared with the tetragonal and cubic YSZ, the BE of the 3d Zr/Y are shifted toward the higher binding energies. This shift increases with the manganese content especially concerning the BE of the Zr 3d core levels. For the all three compounds, the energies of the Zr 3d increase of about 0.7 eV, whereas the energies of the Y 3d increase about 0.55 eV. The spin-orbit splitting between the $d_{5/2}$ and $d_{3/2}$ for Zr and Y was found to be 2.37 and 2.08 respectively and they are in agreement with other studies [24-26].

The magnetic measurements for all compounds were performed as a function of field and temperature. The susceptibility as a function of temperature is shown in Figure 4 (a). All samples display paramagnetic behavior in the whole range of measured temperatures, except the 15% and 20% Mn doped samples which exhibit an anomaly below 47 K coming from the ferrimagnetic transition of the $Mn_3O_4$ impurity phase (47 K) [18]. The inverse of susceptibility for all samples follows a Curie law with Curie–Weiss temperature of $\theta$ = -15.1 K, -23.1 K, -26.0 K and -58 K for 5, 10, 15 and 20% Mn doped compounds respectively. The negative sign of the Curie–Weiss temperature indicates the presence of antiferromagnetic (AFM) interactions in all the samples. The increase of $\theta$ suggests that the AFM interactions are strengthened with the increase of the Mn content which might simply be due to superexchange interactions of close neighbors. The effective paramagnetic moment, $\mu_{eff}$ was determined from the fitted curie constant C (x) by using the relation $\mu_{eff} = \sqrt{3k_B C/(N\mu_0\mu_B^2)}$ with $C(x) = xC_0$ and it was found to be 4.85 $\mu_B$/Mn, 4.60 $\mu_B$/Mn, 4.32 $\mu_B$/Mn, 4.06 $\mu_B$/Mn for 5, 10, 15 and 20 % Mn doped samples respectively. A Mn ion ($Mn^{2+}$, $Mn^{3+}$, $Mn^{4+}$) can exist in two different spin configurations, namely, the high spin state and the low spin state. Then, the maximum paramagnetic moment can be 5.92 $\mu_B$, 4.90 $\mu_B$, 3.87 $\mu_B$ for $Mn^{2+}$, $Mn^{3+}$ and $Mn^{4+}$ respectively in the high spin state. The observed effective paramagnetic moments from M-T measurements are very close to those because of high spin state of $Mn^{3+}$ ions as also confirmed from the XPS measurements. To understand the magnetic property in more detailed, high-field magnetization measurements, $M$ (H), were performed at 3 K (Figure 4 b). The magnetization does not reach to saturation even at the largest applied field of 10 T and no hysteresis was observed up to 15 % Mn doped compounds. This is a general trend expected for a paramagnetic system at moderated temperatures. However, 20 % Mn doped compound was found to exhibit a small hysteresis loop which might be at the cost of $Mn_3O_4$ phase. Thus, the temperature variation of magnetization and high-field magnetic measurements provide clear evidence for a paramagnetic behavior with dominant antiferromagnetic coupling between the manganese atoms, which increase with the Mn concentrations. However for 15% and 20 % of Mn doping, we have observed the ferri-magnetic behavior due to the $Mn_3O_4$ phase. Thus, our experimental data has shown that no ferromagnetic long range order is possible and suggests frustrating effects due to AF couplings.

## IV CONCLUSIONS

To conclude, we have prepared Y-stabilized zirconia, doped with different concentrations of Mn by using solid state method. The samples are found to crystallize into tetragonal symmetry and homogenous up to Mn concentration of 15%, where $Mn_3O_4$ forms at the grain boundaries. The spectral splitting of the Mn 3*s* core-level x-ray photoelectron spectra confirms that Mn ions are in the $Mn^{3+}$ oxidation state. We did not observe any long range ferromagnetic order for *tetragonal* Mn-doped zirconia that were grown *at equilibrium*. Compounds up to 10% of Mn doping are found to be paramagnetic with antiferromagnetic interactions. However, 15% and 20% Mn doped compounds exhibit local ferrimagnetic ordering due to the $Mn_3O_4$ secondary phase. This addresses several important issue like (i) whether the crystal symmetry (tetragonal for present case) play such a crucial role (ii) Since, no ferromagnetism is observed experimentally for cubic Mn-doped zirconia (done by other experimental group) as well as for tetragonal symmetry (our work), theoretical calculation should guide to overcome the discrepancy between experimental and theory. In particular, to allow a direct comparison with our results, it would be of great interest to redo similar calculations using the tetragonal symmetry. If no ferromagnetism is obtained as observed here, one should explain why the symmetry plays such a crucial role.

# LIST OF FIGURES:

**Figure 1:** XRD patterns of $Y_{0.15}(Zr_{0.85-y}Mn_y)O_2$ (y=0, 0.05, 0.10, 0.15 & 0.20) samples. $Y_{0.15}Zr_{0.85}O_2$ and 5% Mn doped compounds were crystallized into tetragonal and monoclinic symmetry. However, higher Mn doped compounds are found to be in tetragonal symmetry. m and t represents the allowed diffraction peak for monoclinic, tetragonal symmetry respectively.

**Figure 2: (a)** SEM micrograph of $Y_{0.15}(Zr_{0.70}Mn_{0.15})O_2$ showing the grain boundary. Elemental color mapping by EDS are shown individually for Zr, Y, Mn, & O in Figures **(b)**, **(c)**, **(d)**, and **(e)** respectively. This indicates that the grain boundary region is poor in Zr & Y but rich in Mn and O elements. **(f)** Variation of lattice parameters (a and c) with Mn doping concentrations.

**Figure 3 :** X-ray photoelectron spectrum of the Mn 2p (a), 3s (b), Zr 3d (c) and Y 3d (d) core levels measured in polycrystalline samples of $Y_{0.15}(Zr_{0.85-y}Mn_y)O_2$. The content of manganese, y, is indicated on the figures. In the figures (a) and (b), Mn 2p and 3s XPS spectra for $MnO_2$ powder are also shown. The arrows indicate the maximum of the peaks positions, obtained from curves fitting.

**Figure 4:** (a) Temperature variation of magnetization for $Y_{0.15}(Zr_{0.85-y}Mn_y)O_2$ (y=0, 0.05, 0.10, 0.15 & 0.20) measured at 0.02 T field. (b) M-H loops recorded for all samples at 3 K.

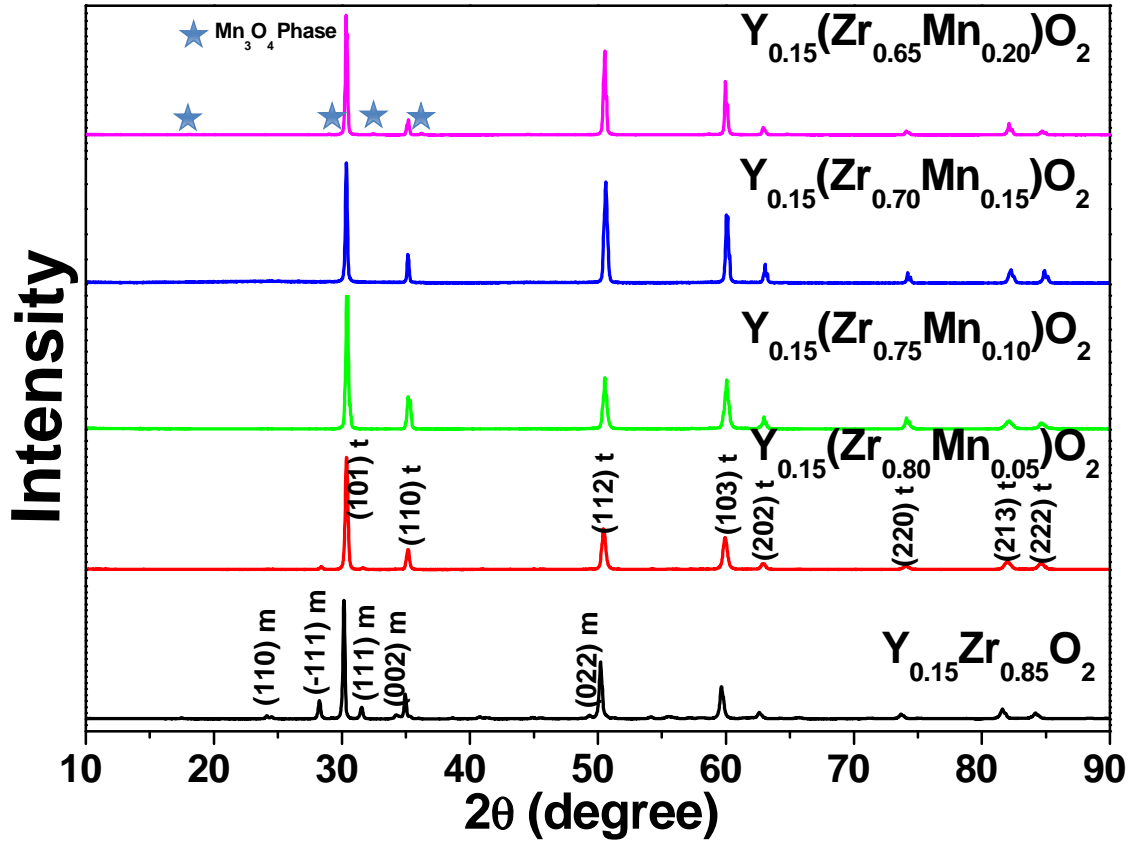

**Figure 1:** XRD patterns of $Y_{0.15}(Zr_{0.85-y}Mn_y)O_2$ (y=0, 0.05, 0.10, 0.15 & 0.20) samples. $Y_{0.15}Zr_{0.85}O_2$ and 5% Mn doped compounds were crystallized into tetragonal and monoclinic symmetry. However, higher Mn doped compounds are found to be in tetragonal symmetry. m and t represents the allowed diffraction peak for monoclinic, tetragonal symmetry respectively.

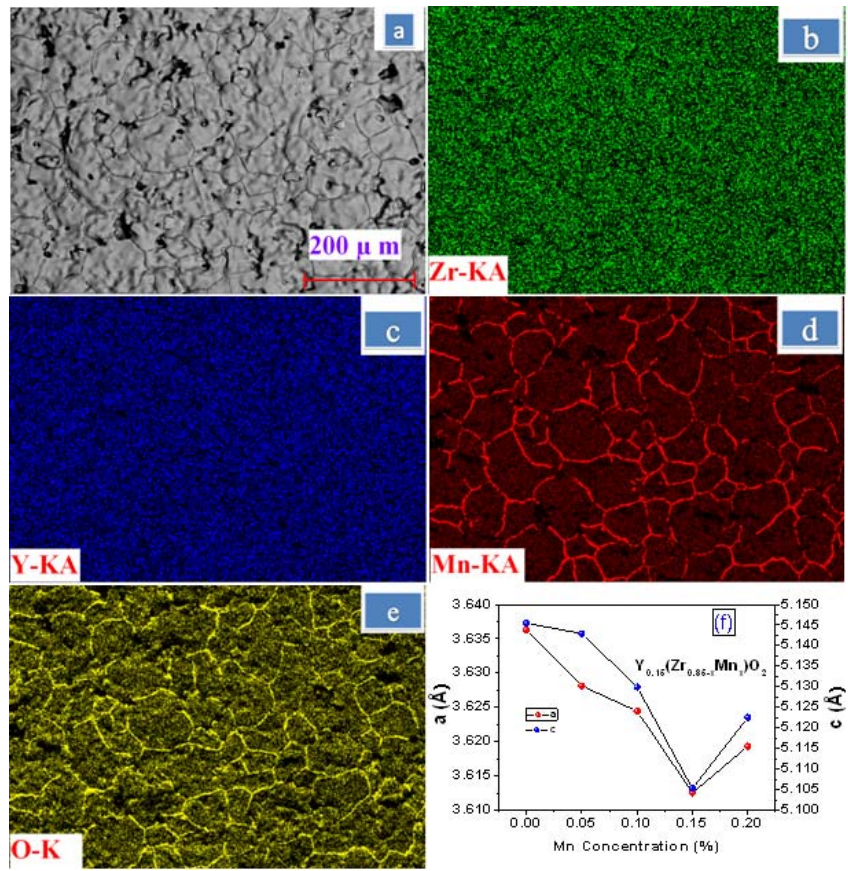

**Figure 2:** **(a)** SEM micrograph of $Y_{0.15}(Zr_{0.70}Mn_{0.15})O_2$ showing the grain boundary. Elemental color mapping by EDS are shown individually for Zr, Y, Mn, & O in Figures **(b)**, **(c)**, **(d)**, and **(e)** respectively. This indicates that the grain boundary region is poor in Zr & Y but rich in Mn and O elements. **(f)** Variation of lattice parameters (a and c) with Mn doping concentrations.

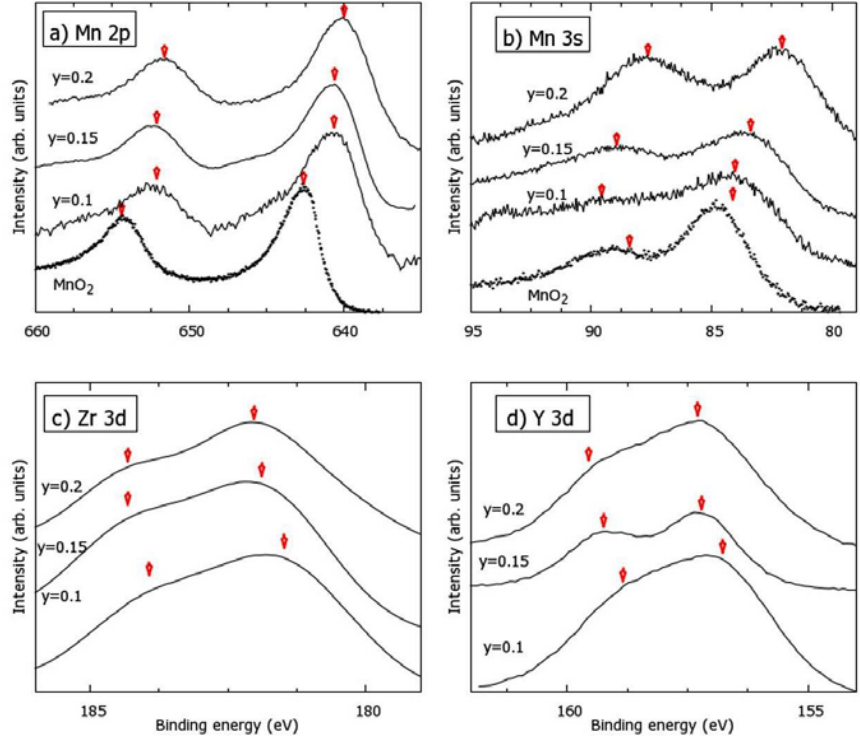

**Figure 3 :** X-ray photoelectron spectrum of the Mn 2p (a), 3s (b), Zr 3d (c) and Y 3d (d) core levels measured in polycrystalline samples of $Y_{0.15}(Zr_{0.85-y}Mn_y)O_2$. The content of manganese, y, is indicated on the figures. In the figures (a) and (b), Mn 2p and 3s XPS spectra for $MnO_2$ powder are also shown. The arrows indicate the maximum of the peaks positions, obtained from curves fitting.

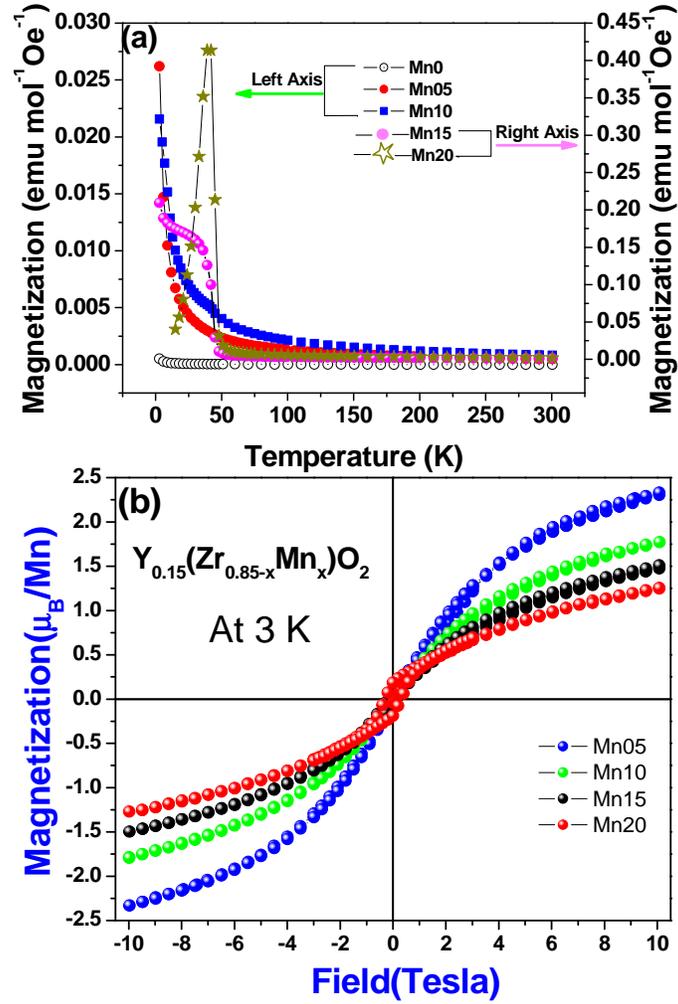

**Figure 4:** (a) Temperature variation of magnetization for $Y_{0.15}(Zr_{0.85-y}Mn_y)O_2$ (y=0, 0.05, 0.10, 0.15 & 0.20) measured at 0.02 T field. (b) M-H loops recorded for all samples at 3 K.

**Table 1:** XPS peaks positions in $Y_{0.15}(Zr_{0.85-y}Mn_y)O_2$ samples obtained from the curves fitting to the XPS spectra shown in the Figure 3. The spin-orbit splitting, $\Delta_{so}$ of the 3d and 2p and the exchange splitting, $\Delta$ of the 3s is deduced from the peak positions of the dominant peaks in the spectra. The Zr and Y 3d peaks positions of the yttria-stabilized zirconia in the cubic and tetragonal crystal structures are also indicated for comparison [24-26].

| Samples | Peak Positions Zr3d (eV) | | Peak positions Y3d (eV) | | Peak positions Mn2p (eV) | | Peak positions Mn3s (eV) | |
|---|---|---|---|---|---|---|---|---|
| | $d_{3/2}$ | $d_{5/2}$ | $d_{3/2}$ | $d_{5/2}$ | $p_{1/2}$ | $p_{3/2}$ | -- | -- |
| Tetragonal YSZ | 184 | 181.6 | 158.3 | 156.2 | -- | | -- | |
| Cubic YSZ | 184.2 | 181.8 | 159.2 | 157.1 | -- | | -- | |
| MnO$_2$ | | | | | 654.3 | 642.6 | 88.406 | 84.406 |
| | -- | | -- | | $\Delta_{so}$=11.53 | | $\Delta$ = 4.28 | |
| y=0.1 | 183.93 | 181.48 | 158.84 | 156.77 | 652.16 | 640.63 | 89.56 | 84.026 |
| | $\Delta_{so}$=2.45 | | $\Delta_{so}$=2.07 | | $\Delta_{so}$=11.53 | | $\Delta$ = 5.55 | |
| y=0.15 | 184.32 | 181.89 | 159.24 | 157.21 | 652.12 | 640.59 | 88.94 | 83.38 |
| | $\Delta_{so}$=2.37 | | $\Delta_{so}$=2.03 | | $\Delta_{so}$=11.53 | | $\Delta$ = 5.56 | |
| y=0.2 | 184.32 | 182.03 | 159.55 | 157.39 | 651.62 | 639.98 | 87.64 | 82.08 |

| | $\Delta_{so}$=2.3 | $\Delta_{so}$=2.16 | $\Delta_{so}$=11.64 | $\Delta$ = 5.57 |